\documentclass[aps,prc,twocolumn,superscriptaddress,showpacs,10pt,dvipsnames]{revtex4-1}

\usepackage[utf8]{inputenc}
\usepackage{amsfonts}
\usepackage{amsmath}
\usepackage{amssymb}
\usepackage{bm}
\usepackage[pdftex]{graphicx}
\usepackage{xcolor}
\usepackage[normalem]{ulem}

\begin{document}

\newcommand{\tj}[6]{ \begin{pmatrix}
  #1 & #2 & #3 \\
  #4 & #5 & #6
 \end{pmatrix}}


\title{Radiative capture reaction for $^{17}$Ne formation within a full three-body model} 




\author{J. Casal}
\email{jcasal@us.es}
\affiliation{Departamento de F\'{\i}sica At\'omica, Molecular y Nuclear,
  Facultad de F\'{\i}sica, Universidad de Sevilla, Apartado 1065, E-41080
  Sevilla, Spain} 
\affiliation{European Centre for Theoretical Studies in Nuclear Physics and Related Areas (ECT$^*$) and Fondazione Bruno Kessler, Villa Tambosi, Strada delle Tabarelle 286, I-38123 Villazzano (TN), Italy}
\author{E. Garrido}
\email{e.garrido@csic.es}
\affiliation{Instituto de Estructura de la Materia, IEM-CSIC, Serrano 123, E-28006 Madrid, Spain}
\author{R. de Diego}
\affiliation{Centro de Ciências e Tecnologias Nucleares, Instituto Superior Técnico,
Universidade de Lisboa, Estrada Nacional 10 (Km 139,7), P-2695-066 Bobadela LRS, Portugal}
\author{J. M. Arias}
\affiliation{Departamento de F\'{\i}sica At\'omica, Molecular y Nuclear,
  Facultad de F\'{\i}sica, Universidad de Sevilla, Apartado 1065, E-41080
  Sevilla, Spain}
\author{M. Rodr\'{\i}guez-Gallardo}
\affiliation{Departamento de F\'{\i}sica At\'omica, Molecular y Nuclear,
  Facultad de F\'{\i}sica, Universidad de Sevilla, Apartado 1065, E-41080
  Sevilla, Spain}


\date{\today}

\begin{abstract}
 \begin{description}
  \item[Background] The breakout from the hot Carbon-Nitrogen-Oxigen (CNO) cycles can trigger the rp-process in type I x-ray bursts. In this environment, a competition between $^{15}\text{O}(\alpha,\gamma){^{19}\text{Ne}}$ and the two-proton capture reaction $^{15}\text{O}(2p,\gamma){^{17}\text{Ne}}$ is expected.
  \item[Purpose] Determine the three-body radiative capture reaction rate for ${^{17}\text{Ne}}$ formation including sequential and direct, resonant and non-resonant contributions on an equal footing.
  \item[Method] Two different discretization methods have been applied to generate $^{17}$Ne states in a full three-body model: the analytical transformed harmonic oscillator method and the hyperspherical adiabatic expansion method. The binary $p$--$^{15}$O interaction has been adjusted to reproduce the known spectrum of the unbound $^{16}$F nucleus. The dominant $E1$ contributions to the $^{15}\text{O}(2p,\gamma){^{17}\text{Ne}}$ reaction rate have been calculated from the inverse photodissociation process.
  \item[Results] Three-body calculations provide a reliable description of $^{17}$Ne states. The agreement with the available experimental data on $^{17}$Ne is discussed. It is shown that the $^{15}\text{O}(2p,\gamma){^{17}\text{Ne}}$ reaction rates computed within the two methods agree in a broad range of temperatures. The present calculations are compared with a previous theoretical estimation of the reaction rate.
  \item[Conclusions] It is found that the full three-body model provides a reaction rate several orders of magnitude larger than the only previous estimation. The implications for the rp-process in type I x-ray bursts should be investigated.
 \end{description} 
\end{abstract}


\maketitle


\section{Introduction}
Nucleosynthesis in explosive scenarios at the final stages of stellar evolution follows reaction paths involving exotic nuclei~\cite{Langanke01}. Explosive H and He burning at high temperatures can trigger the rp-process in type I x-ray bursts~\cite{Schatz98}. These are binary systems consisting of a red giant and a neutron star, where the neutron star accretes H-rich matter from the companion star. The proton flux is heated and compressed, leading the rp-process to potentially populate nuclides off the hot Carbon-Nitrogen-Oxigen (CNO) cycle, i.e.~Ne, F, and Na via breakout reactions on waiting-point nuclei~\cite{Wiescher99}. These reactions rapidly converts the light-element fuel into heavier, proton-rich nuclei~\cite{CIliadisNPStars}. The balance between the slow, $\beta$-limited CNO cycles and the rp-process controls the trigger conditions of the x-ray burst~\cite{Wiescher10}. Among the relevant reactions, $^{15}\text{O}(\alpha,\gamma){^{19}\text{Ne}}$ and $^{18}\text{Ne}(\alpha,p){^{21}\text{Na}}$ are the most representative~\cite{Schatz98,Wiescher10}. But, as an alternative, the two-proton capture reaction $^{15}\text{O}(2p,\gamma){^{17}\text{Ne}}$ may also play a relevant role~\cite{Gorres95,Grigorenko05}. The resonant~\cite{Grigorenko05} and non-resonant~\cite{Grigorenko06} capture processes for the production of $^{17}$Ne have been studied theoretically by Grigorenko \textit{et al.}, showing the relevance of the three-body direct capture compared to sequential estimations~\cite{Gorres95}. 

The $^{17}$Ne nucleus can be studied within an $^{15}\text{O}+p+p$ three-body model. Since the proton capture on $^{15}$O leads to an unbound $^{16}$F system, $^{17}$Ne presents a Borromean structure. Besides the relevance of $^{17}$Ne for the rp-process in x-ray bursts, this nucleus has attracted special interest over the past years, as it is the most promising known candidate to present a two proton halo. Despite the remarkable efforts to address the structure of $^{17}$Ne, controversy still exists~\cite{Garrido04,Tanaka10}. The halo nature of $^{17}$Ne has not yet been confirmed.

Recently, we have presented three-body calculations regarding the formation of Borromean nuclei within a full three-body model~\cite{RdDiego10,JCasal13,JCasal14}, treating resonant and non-resonant, sequential and direct contributions on an equal footing. This is a fundamental difference compared to the results in Refs.~\cite{Grigorenko05,Grigorenko06} for $^{17}$Ne, in which the resonant and non-resonant contributions to the reaction rate were computed separately. For weakly-bound systems, such as $^{6}$He or $^9$Be, their small separation energy implies large breakup probability in scattering processes. This can be understood as an excitation of the nucleus to unbound states that form a continuum of energies~\cite{Dasso04reg}. On the other hand, the synthesis of nuclei in stellar environments can be described as a decay from an unbound state of several particles that fuse together, producing a bound system~\cite{RdDiego10}. Both processes demand a reasonable treatment of continuum states. 

In general, the treatment of continuum states is a difficult task, since their asymptotic behavior for three-body systems comprising several charged particles is not known in general. A possibility consists of using the so-called discretization methods~\cite{Austern87,Rasoanaivo89}. These methods replace the actual continuum by a finite set of normalizable states, i.e., a discrete basis that can be truncated to a relatively small number of states providing a reasonable description of the system. 

In this work, we address the $^{15}\text{O}(2p,\gamma){^{17}\text{Ne}}$ reaction rate. The only available estimations of the radiative capture reaction rate for $^{17}$Ne formation are those in Refs.~\cite{Grigorenko05,Grigorenko06}. The main objective of this work is to present a comprehensive three-body calculation that provides a unified description of the sequential and direct, resonant and non-resonant capture. For this purpose, we use two different discretization methods to describe $^{17}$Ne states: 
i) The analytical transformed harmonic oscillator (THO) method within the hyperspherical harmonic (HH) framework~\cite{MRoGa05,JCasal13}. 
ii) The hyperspherical adiabatic (HA) expansion method~\cite{Nielsen01} with a box boundary condition.
In both approaches, the negative-energy solutions describe the bound states of the system, while positive-energy solutions are taken as a discrete representation of the continuum. These methods can be applied to a general three-body system comprising any number of charged clusters.

This paper is structured as follows. In Sec.~\ref{sec:theory}, the three-body formalism is presented. In Sec.~\ref{sec:application}, the method is applied to describe the structure of $^{17}$Ne, and the rate of the radiative capture reaction $^{15}\text{O}(2p,\gamma){^{17}\text{Ne}}$ is obtained. Finally, Sec.~\ref{sec:conclusions} summarizes the main conclusions of this work. 

\section{Formalism}
\label{sec:theory}

\begin{figure}
\centering
 \includegraphics[width=\linewidth]{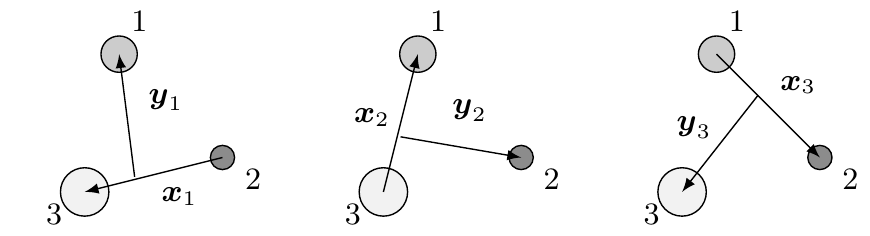}
 \caption{The three sets of scaled Jacobi coordinates.}
 \label{fig:sets}
\end{figure}

Three-body systems can be described using Jacobi coordinates $\{\boldsymbol{x}_k,\boldsymbol{y}_k\}$, where the label $k$ indicates one of the three coordinate sets in Fig.~\ref{fig:sets}. The variable $\boldsymbol{x}_k$ is proportional to the relative coordinate between two particles and $\boldsymbol{y}_k$ is proportional to the distance from the center of mass of the $x$ subsystem to the third particle, both with a scaling factor depending on their masses~\cite{MRoGa05}. As in Ref.~\cite{JCasal15}, we use the notation in which, for example, the Jacobi-1 system corresponds to the system where the particles (2,3) are related by the coordinate $\boldsymbol{x}_1$. From Jacobi coordinates, the hyperspherical coordinates $\{\rho,\alpha_k,\widehat{x}_k,\widehat{y}_k\}$ are introduced. Here, the hyper-radius $(\rho)$ and the hyperangle $(\alpha_k)$ are given by
\begin{align}
\rho = & \sqrt{x_k^2 + y_k^2}, \\
\alpha_k = & \tan\left(\frac{x_k}{y_k}\right),
\label{eq:jachyp}
\end{align}
and $\{\widehat{x}_k,\widehat{y}_k\}$ are the two-dimensional angular variables related to $\{\boldsymbol{x}_k,\boldsymbol{y}_k\}$. Note that, while the hyperangle depends on $k$, the hyper-radius does not.

We consider the radiative capture reaction rate of three particles $(abc)$, into a bound nucleus $A$ of binding energy $|\varepsilon_B|$, i.e., $a+b+c\to A+\gamma$. The energy-averaged reaction rate for such process can be obtained from the inverse photodissociation process and is given as a function of the temperature by the expression~\cite{RdDiego10,JCasal13}
\begin{equation}
\langle R_{abc}(\varepsilon)\rangle (T) = \mathcal{C}(T) \int_{|\varepsilon_B|}^{\infty} d\varepsilon_\gamma~\varepsilon_\gamma^2 \sigma_\gamma(\varepsilon_\gamma)e^{\frac{-\varepsilon_\gamma}{k_BT}},
\label{eq:rate1}
\end{equation}
where $\varepsilon=\varepsilon_\gamma+\varepsilon_B$ is the initial three-body kinetic energy, $\varepsilon_\gamma$ is the energy of the photon emitted, $\varepsilon_{B}$ is the ground-state energy, $\sigma_\gamma(\varepsilon_\gamma)$ is the photodissociation cross section of $A$, and $\mathcal{C}(T)$ is a temperature-dependent constant given by
\begin{equation}
\mathcal{C}(T) = \nu!\frac{\hbar^3}{c^2}\frac{8\pi}{\left(a_xa_y\right)^{3/2}}\frac{g_A}{g_ag_bg_c}\frac{e^{\frac{|\varepsilon_B|}{k_BT}}}{\left(k_BT\right)^3}.
\end{equation}
Here, $g_i$ are the spin degeneracies of the particles, $\nu$ is the number of identical particles in the three-body system, and $a_x$, $a_y$ are the reduced masses of the subsystems related to Jacobi coordinates $\{\boldsymbol{x},\boldsymbol{y}\}$. Note that the reaction rate in Eq.~(\ref{eq:rate1}) could be computed, provided the experimental photodissociation cross section is known for the compound nucleus. However, direct photodissociation measurements can be done only for stable nuclei, e.g., $^{12}$C~\cite{Gai11}, sometimes with important discrepancies among different experiments, e.g., $^{9}$Be~\cite{Sumiyoshi02,Arnold12}. Thus, for reactions involving unstable nuclei, this technique is not feasible. Recently, an alternative procedure was proposed to obtain three-body radiative capture reaction rates from experimental information on inclusive break-up reactions at low energies~\cite{JCasal16R}. No such data are available in the literature for $^{17}$Ne, so theoretical models to describe its structure are in order.

The photodissociation cross section in Eq.~(\ref{eq:rate1}) can be expanded into electric and magnetic multipoles~\cite{RdDiego10,Forseen03}
\begin{equation}
\sigma_\gamma^{(\mathcal{O}\lambda)}(\varepsilon_\gamma)=\frac{(2\pi)^3 (\lambda+1)}
{\lambda[(2\lambda+1)!!]^2}\left(\frac{\varepsilon_\gamma}{\hbar
c}\right)^{2\lambda-1}\frac{dB(\mathcal{O}\lambda)}{d\varepsilon},
\label{eq:xsection}
\end{equation}
which are related to the transition probability distributions $dB(\mathcal{O}\lambda)/d\varepsilon$, for $\mathcal{O}=E, M$. The integral in Eq.~(\ref{eq:rate1}) is very sensitive to the behavior of the transition probability distributions at low energies, thus requiring a detailed description of the low-energy continuum. 

In a discrete representation, the reduced transition probability between states of the system is defined, following the notation of Brink and Satchler~\cite{BrinkSatchler}, as
\begin{eqnarray}
 \nonumber B(\mathcal{O}\lambda)_{nj,n'j'} & \equiv & B(\mathcal{O}\lambda;nj\rightarrow n'j') \\
 & =& |\langle nj\|\widehat{\mathcal{O}}_\lambda\|n'j'\rangle|^2\left(\frac{2\lambda+1}{4\pi}\right),
\label{eq:BE}
\end{eqnarray}
where $\widehat{\mathcal{O}}_{\lambda M_\lambda}$ is the electric or magnetic multipole operator of order $\lambda$. In the case of electric transitions, the multipole operator can be written in the Jacobi-$k$ set as
\begin{equation}
\widehat{\mathcal{O}}_{\lambda M_{\lambda}}(\boldsymbol{x}_k,\boldsymbol{y}_k)=\left(\frac{4\pi}{2\lambda+1}\right)^{1/2}\sum_{q=1}^3 Z_q~e~r_q^\lambda Y_{\lambda M_{\lambda}}(\widehat{r}_q),
\label{eq:Qop}
\end{equation}
where $Z_q$ is the atomic number of the particle $q$, $e$ is the electron charge, and $\boldsymbol{r}_q$ is the position of particle $q$ with respect to the center of mass of the system, which in the 
Jacobi-$q$ system is given by~\cite{Nielsen01} 
\begin{equation}
 \boldsymbol{r}_q = \sqrt{\frac{m}{m_q}\frac{\left(M_T-m_q\right)}{M_T}}\boldsymbol{y}_q.
 \label{eq:relpos}
\end{equation}
Here $m$ is a normalization mass, taken as the atomic mass unit, and $M_T$ is the total mass of the system. We describe the system in a preferred Jacobi set, $k$; however, the expression for the electric multipole operator given by Eq.~(\ref{eq:BE}) can be easily expressed, in general, using different Jacobi systems. The relation between  harmonic polynomials in different Jacobi sets is given by the expression~\cite{RdDiego08}
\begin{align}
 y_q^\lambda Y_{\lambda M_\lambda}\left(\widehat{y}_q\right)  = & \sum_{l=0}^{\lambda}\left(-1\right)^\lambda x_k^{\lambda-l}\left(\sin\varphi_{qk}\right)^{\lambda-l} y_k^l \left(\cos\varphi_{qk}\right)^l \nonumber\\
& \times \sqrt{\frac{4\pi\left( 2\lambda+1\right)!}{\left( 2l+1\right)!\left( 2\lambda-2l+1\right)!}} \nonumber \\
& \times \left[Y_{\lambda-l}\left(\widehat{x}_k\right) \otimes Y_{l}\left( \widehat{y}_k\right) \right]^{\lambda M_\lambda},
 \label{eq:harmonicp}
 \end{align}
with
 \begin{equation}
 \tan\varphi_{qk} = \left( -1\right)^P\sqrt{\frac{m_p M_T}{m_q m_k}},
 \label{eq:phi}
 \end{equation}
depending on the mass of the particles and the parity $(-1)^P$ of the permutation $P$ of $\{k,p,q\}$. The identity transformation is given by $\varphi_{kk}=\pi$. 
Using Eq.~(\ref{eq:harmonicp}) we can rewrite the harmonic polynomial for each particle $q$, as a function of the Jacobi coordinates in the preferred Jacobi system $k$. 
Details regarding the computations of the matrix elements of $\mathcal{O}_{\lambda M_\lambda}$ can be found, for instance, in Refs.~\cite{JCasal13,JCasal14}. Note that the ket $|nj\mu\rangle$ represents the wave function of the system with angular momentum $j$ and projection $\mu$, with $n$ being a label which enumerates the states. These states can be obtained using different discretization methods. In the following sections, the two approaches used in this work are schematically presented.

\subsection{The THO method within the HH framework}
\label{sec:THO}
In the hyperspherical harmonic (HH) formalism, the eigenstates of the system in a fixed Jacobi set can be expanded as 
\begin{equation}
 \Psi^{nj\mu}(\rho,\Omega)=\frac{1}{\rho^{5/2}}\sum_{\beta} \chi_{n\beta}^{j\mu}(\rho)\mathcal{Y}_{\beta j\mu}(\Omega),
 \label{eq:wf}
\end{equation}
where $\Omega\equiv\{\alpha,\widehat{x},\widehat{y}\}$ is introduced for the angular dependence and $\beta\equiv\{K,l_x,l_y,l,S_x,j_{ab}\}$ is a set of quantum numbers we call channel. In this set, $K$ is the hypermomentum, $l_x$ and $l_y$ are the orbital angular momenta associated with the Jacobi coordinates $\boldsymbol{x}$ and $\boldsymbol{y}$, respectively, $l$  is the total orbital angular momentum ($\boldsymbol{l}=\boldsymbol{l_x}+\boldsymbol{l_y}$), $S_x$ is the spin of the particles related by the coordinate $\boldsymbol{x}$, and $j_{ab}$ results from the coupling $\boldsymbol{j_{ab}}=\boldsymbol{l}+\boldsymbol{S_x}$. If we denote by $I$  the spin of the third particle, that we assume to be fixed, the total angular momentum $j$ is $\boldsymbol{j}=\boldsymbol{j_{ab}} + \boldsymbol{I}$. Notice that, for simplicity, the label $k$ has been omitted. The functions $\mathcal{Y}_{\beta j\mu}(\Omega)$ are states of good total angular momentum, expanded in hyperspherical harmonics~\cite{Zhukov93}.

The radial functions $\chi_{n\beta}^{j\mu}(\rho)$ in Eq.~(\ref{eq:wf}) can be obtained using the pseudo-state (PS) method~\cite{SiegertPS}, which consist in diagonalizing the Hamiltonian in a complete set of square-integrable functions. For this purpose, a variety of bases have been proposed for two-body~\cite{HaziTaylor70,Matsumoto03,MRoGa04,AMoro09} and three-body systems~\cite{Desc03,Matsumoto04,MRoGa05,JCasal13}. In this work, as in Refs.~\cite{JCasal13,JCasal14,JCasal15}, we use the analytical transformed harmonic oscillator (THO) basis, so we can write
\begin{equation}
 \chi_{n\beta}^{j\mu}(\rho) = \sum_{i} C_n^{i\beta j} U_{i\beta}^\text{THO}(\rho),
 \label{eq:expandTHO}
\end{equation}
where $i$ denotes the hyperradial excitation and $C_n^{i\beta j}$ are just the diagonalization coefficients. Therefore, Eq.~(\ref{eq:wf}) involves infinite sums over $\beta$ and $i$. However, calculations are typically truncated at maximum hypermomentum $K_{max}$ and $i_{max}$ hyperradial excitations in each channel. These parameters have to be large enough to provide converged results. 

The THO basis functions in Eq.~(\ref{eq:expandTHO}) are obtained from the harmonic oscillator (HO) functions using a local scale transformation $s(\rho)$,
\begin{equation}
  U_{i\beta}^{\text{THO}}(\rho)=\sqrt{\frac{ds}{d\rho}}U_{iK}^{\text{HO}}[s(\rho)].
\label{eq:R}
\end{equation}
This transformation keeps the simplicity of the HO functions, but converts their Gaussian asymptotic behavior into an exponential one. This provides a suitable representation of bound and resonant states to calculate structure and scattering observables. We use the analytical form proposed by Karataglidis \textsl{et al.}~\cite{Karataglidis},
\begin{equation}
s(\rho) = \frac{1}{\sqrt{2}b}\left[\frac{1}{\left(\frac{1}{\rho}\right)^{4} +
\left(\frac{1}{\gamma\sqrt{\rho}}\right)^4}\right]^{\frac{1}{4}},
\label{eq:LST}
\end{equation}
depending on the parameters $\gamma$ and $b$. Note that the THO hyperradial wave functions depend, in general, on all the quantum numbers included in a channel $\beta$, although the HO hyperradial wave functions only depend on the hypermomentum $K$. The most interesting feature of the analytical THO method is that the ratio $\gamma/b$ governs the asymptotic behavior of the basis functions and controls the density of PSs as a function of the energy. This allows us to select an optimal basis depending on the system or observable under study~\cite{JCasal13}.

\subsection{The HA expansion method in a box}
\label{sec:adiabatic}

Following Ref.~\cite{Nielsen01}, we give here a brief sketch of the hyperspherical adiabatic (HA) expansion method.
Using the hyperspherical coordinates introduced in Sec.~\ref{sec:theory}, the
three-body Hamiltonian $\hat{\cal H}$ takes the form:
\begin{equation}
\hat{\cal H} =  -\frac{\hbar^2}{2 m} \hat{T}_\rho + \frac{\hbar^2}{2 m \rho^2}\hat{\Lambda}^2
+ V(\rho,\Omega)
 =  -\frac{\hbar^2}{2 m} \hat{T}_\rho + \hat{{\cal H}}_\Omega ,
\label{eq1}
\end{equation}
where $\hat{T}_\rho=\frac{\partial^2}{\partial\rho^2}+\frac{5}{\rho}\frac{\partial}{\partial\rho}$ 
is the hyperradial kinetic energy operator, and ${\cal H}_\Omega$  contains the whole dependence on 
the hyperangles. In the expression above $\hat{\Lambda}^2$ is the hyperangular
operator, $V(\rho,\Omega)=\sum_i V_i(x_i)$ is the sum of the three two-body potentials,
and $m$ is the normalization mass used to define the Jacobi coordinates.

In the HA expansion method the Schr\"{o}dinger equation $(\hat{\cal H}-E)\Psi=0$ is solved in two steps.
In the first one, for given three-body quantum numbers $\{n,j,\mu\}$, the angular part is solved for 
a set of fixed values of $\rho$. This amounts to solving the eigenvalue problem 
\begin{equation}
\hat{{\cal H}}_\Omega \Phi_\nu^{j\mu}(\rho,\Omega)=\frac{\hbar^2}{2 m} \frac{1}{\rho^2}
\lambda_\nu^j(\rho) \Phi_\nu^{j\mu}(\rho,\Omega)
\label{eq2}
\end{equation}
for each $\rho$, which is treated as a parameter. This eigenvalue problem is solved after expansion of the
angular functions $\Phi_\nu^{j\mu}(\rho,\Omega)$ in terms of the $\{ {\cal Y}_{\beta j\mu} \}$ functions introduced in Eq.(\ref{eq:wf}).

The angular functions $\{\Phi_\nu^{j\mu}(\rho, \Omega)\}$ form a complete orthonormal basis (HA basis)
for each value of $\rho$. This basis is now used to expand the full three-body wave function, 
which, instead of by Eq.(\ref{eq:wf}), is now given by:
\begin{equation}
\Psi^{n j \mu}(\rho,\Omega) = \frac{1}{\rho^{5/2}}\sum_{\nu=1}^\infty 
                     f_\nu^{nj}(\rho) \Phi_\nu^{j\mu}(\rho,\Omega).
\label{eq3}
\end{equation}
Obviously the summation above has to be truncated, and only a finite number of adiabatic terms are
included in the calculation. Tipically, no more than 10 adiabatic terms are enough to get convergence.

In a second step, the radial wave functions $f_\nu^{n j}(\rho)$ in the expansion (\ref{eq3}) 
are obtained after solving the following coupled set of radial equations:
\begin{eqnarray}
\lefteqn{ \hspace*{-1cm}
\left(
 -\frac{d^2}{d\rho^2} + \frac{1}{\rho^2}
\left( \lambda_\nu^j(\rho)+\frac{15}{4} \right) - \frac{2m\varepsilon}{\hbar^2}\right) f_\nu^{nj}(\rho)=
        } \nonumber \\ & & \hspace*{1cm}
\sum_{\nu'} 
\left(2P_{\nu\nu'}^j(\rho)\frac{\partial}{\partial \rho}+Q_{\nu\nu'}^j(\rho) \right) f_{\nu'}^{nj}(\rho),
\label{eq5}
\end{eqnarray}
where $\varepsilon$ is the three-body energy, and the eigenfunctions $\lambda_\nu^j$ of the hyperangular
Hamiltonian ${\cal H}_\Omega$, Eq.(\ref{eq2}),  enter as effective potentials. Finally, 
the coupling terms $P_{\nu\nu'}^j$ and $Q_{\nu\nu'}^j$ take the form:
\begin{eqnarray}
& & P_{\nu \nu'}^j(\rho)=\langle \Phi_\nu^{j\mu}(\rho,\Omega) \Big|\frac{\partial}{\partial \rho} \Big| 
                           \Phi_{\nu^\prime}^{j\mu}(\rho,\Omega) \rangle_\Omega  \nonumber \\
& & Q_{\nu \nu'}^j(\rho)=\langle \Phi_\nu^{j\mu}(\rho,\Omega) \Big|\frac{\partial^2}{\partial \rho^2} \Big| 
                           \Phi_{\nu^\prime}^{j\mu}(\rho,\Omega) \rangle_\Omega, 
\label{coup}
\end{eqnarray}
where $\langle \rangle_\Omega$ represents integration over the five hyperangles only.

When using the HA expansion method the continuum spectrum will be discretized after solving
the set of radial equations (\ref{eq5}) by imposing a box boundary condition, i.e., the radial
functions $f_\nu^{nj}(\rho)$ are imposed to be zero for some large value of the hyperradius $\rho_{max}$.
This procedure immediately leads to a set of discrete continuum states which are formally treated 
as bound states, and therefore they are just normalized to 1 inside the box.
As shown in Ref.~\cite{Garrido15}, the discrete energy spectrum constructed in this way is not uniformly 
distributed. Instead, the discrete continuum energies appear in groups of states, almost degenerate,
each of them containing as many states as adiabatic terms included in the expansion (\ref{eq3}). 
Eventually, for $\rho_{max}=\infty$ these states are completely degenerate, and they correspond
to all the possible incoming channels for a given energy. In other words,  
the discrete continuum states keep the full information about the three-body state. All the possible 
incoming and outgoing channels are actually taken into account, and therefore the information contained in
the $S$ matrix is fully preserved. Furthermore, as shown in Ref.~\cite{Garrido14}, for a sufficiently large size of the box,
this discretization procedure is equivalent to normalizing the continuum wave functions
matching them to the correct asymptotic behavior. Therefore, this discretization procedure
can be safely used in those cases, like the three-body Coulomb problem,  where the asymptotic 
form of the wave functions is not known analytically. This argument is valid also in the THO case 
for a sufficiently large basis, which is equivalent to a sufficiently large box in the HA case.

\section{Application to $^{17}$Ne}
\label{sec:application}
\begin{figure}[t]
\centering
 \includegraphics[width=0.5\linewidth]{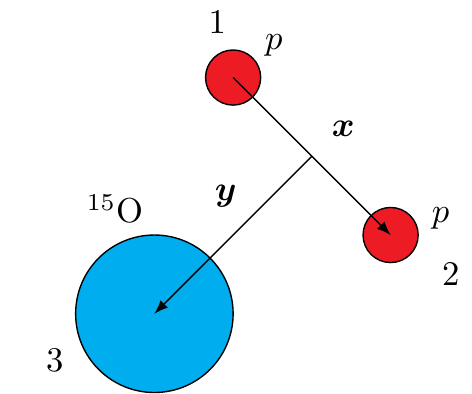}
 \caption{(Color online) The Jacobi-$T$ system used to describe the $^{17}$Ne nucleus.}
 \label{fig:JacobiT17ne}
\end{figure}

In the THO method the $^{17}$Ne nucleus is described in the Jacobi-$T$ system, as shown in Fig.~\ref{fig:JacobiT17ne}, where the two identical protons are related by the coordinate $\boldsymbol{x}$. This choice enables the proper treatment of the Pauli principle by removing the corresponding components of the wave functions (\ref{eq:wf}) that would disappear under full antisymmetrization. When the HA expansion method is used, the Faddeev
equations are solved (see Ref.~\cite{Nielsen01}), in such a way that all the three possible Jacobi sets are
equally treated. In this case the Pauli forbidden states are removed by excluding from the calculation
the adiabatic terms in the expansion (\ref{eq3}) associated to the Pauli forbidden states \cite{Garrido97}.

A three-body description of the Borromean nucleus $^{17}$Ne deals with the complication that the corresponding $^{15}$O core has non-vanishing spin. Core excitations could play a role in describing the structure and dynamics of $^{17}$Ne. However, the lowest excited states in $^{15}$O occur at relatively high energies compared to the first excited states in $^{17}$Ne~\cite{Ajzenberg91,Guimaraes98,Chromik02}. Therefore, the assumption of a structureless core with fixed spin 1/2$^-$ seems to be a reliable picture. As in previous studies about the structure of $^{17}$Ne~\cite{Garrido04}, we neglect core excitations, although their effect on structure and reaction observables needs to be further investigated.

\begin{figure}
\includegraphics[width=\linewidth]{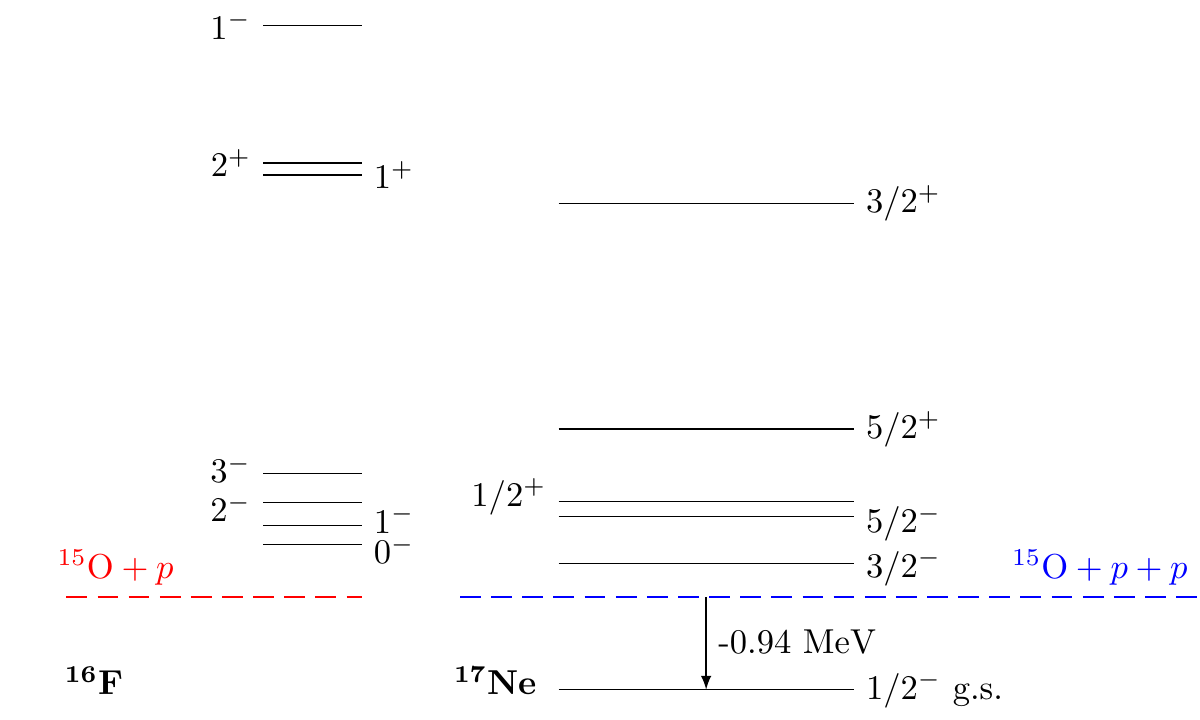}
 \caption{(Color online) Low-lying states of $^{17}$Ne~\cite{Guimaraes98,Chromik02} and the $^{16}$F ($p$ + $^{15}$O)~\cite{Ajzenberg86} subsystem. The energies are given with respect to the $2p$ and $p$ thresholds, respectively.}
 \label{fig:spectrum17Ne}
\end{figure}

We use in the model Hamiltonian for $^{17}$Ne the $p$-$p$ GPT potential~\cite{GPT}, which includes central, spin-orbit, tensor and Coulomb terms. For the $p$-$^{15}$O interaction, some prescription is needed to fit the potentials. We adjust an $l$-dependent interaction with central, spin-orbit and spin-spin terms,
\begin{equation}
V_{p-core}^{(l)} (r) = V_c^{(l)}(r) + \boldsymbol{s}_p\cdot\boldsymbol{l}_x V_{so}(r) + \boldsymbol{s}_p\cdot\boldsymbol{s}_{core} V_{ss}^{(l)}(r),
\label{eq:pot16F}
\end{equation}
to fit the known resonances of the unbound system $^{16}$F. The lowest states in $^{16}$F are shown together with the $^{17}$Ne states in Fig.~\ref{fig:spectrum17Ne}. 
The available experimental data~\cite{Ajzenberg86} on these states are shown in Table~\ref{tab:16F}. In Eq.~(\ref{eq:pot16F}), the form factors for the central ($V_c^{(l)}$) and spin-spin ($V_{ss}^{(l)}$) terms are taken as Woods-Saxon functions, $V^{(l)}(r) = v^{(l)}/[1+\exp(r-b)/a]$, while the spin-orbit potential ($V_{so})$ for the proton is chosen to have a Woods-Saxon derivative form. These potentials have the same radius, $b=3.13$ fm, and the same diffuseness, $a=0.67$ fm. The corresponding $l$-dependent strengths are shown in Table~\ref{tab:str}. This potential, together with a hard-sphere Coulomb interaction with a Coulomb radius of $r_{\rm Coul}=3.13$ fm, provides a good agreement with the experimental energies of the two-body $^{16}$F resonances and is consistent with the results in Ref.~\cite{Garrido04}. Details regarding the calculation of the potential matrix elements for three-body systems can be found, for instance, in Ref.~\cite{IJThompson04}.

\begin{table}
\centering
\begin{tabular}{clccl}
\toprule
$j^\pi$ & $(E_R,\Gamma)$ (MeV) & & $j^\pi$ & $(E_R,\Gamma)$ (MeV)\\
\colrule
0$^-$ & (0.535, 0.040) & & 1$^+$ & (4.29, $<$ 0.040) \\
1$^-$ & (0.728, $<$ 0.040) & & 2$^+$ & (4.41, $<$ 0.020) \\
2$^-$ & (0.959, 0.040) & & 1$^-$ & (5.81, --) \\
3$^-$ & (1.256, $<$ 0.015) & & 2$^-$ & -- \\
\botrule
\end{tabular}
\caption{Experimental two-body spectrum for $^{16}$F~\cite{Ajzenberg86}. The values are given as the resonance energy and the corresponding width, $(E_R,\Gamma)$.}
\label{tab:16F}
\end{table}

\begin{table}
\centering
\begin{tabular}{lcc}
\toprule
$l$ & $v_c^{(l)}$ (MeV) & $v_{ss}^{(l)}$ (MeV)\\
\colrule
0 & -50.0 & 0.7 \\
1 & -11.0 & 1.0 \\
2 & -48.4 & 2.0 \\
\botrule
\end{tabular}
\caption{Strengths of the central ($v_c^{(l)}$) and spin-spin ($v_{ss}^{(l)}$) Woods-Saxon potentials in Eq.~(\ref{eq:pot16F}) as a function of the relative $p$--$^{15}$O angular momentum $l$. The spin-orbit strength is fixed to $v_{so}=-30$ MeV~fm$^2$ for $l=1,2$.}
\label{tab:str}
\end{table}

The preceeding $p$-$^{15}$O potential presents unphysical bound states that correspond to the proton $s_{1/2}$ states occupied in the $^{15}$O core. The Pauli principle has to be taken into account by forbidding these two-body states within three-body calculations. There are different prescriptions available in the literature to address this problem~\cite{IJThompson00}. In this work, we use the adiabatic projection method~\cite{Garrido97} to eliminate the Pauli forbidden states. In addition to the binary interactions, it is customary to include also a simple hyperradial three-body force to adjust the energies of the known three-body states to their experimental positions~\cite{Garrido04,IJThompson04,MRoGa05,JCasal13}. In this work, we use a Gaussian form,
\begin{equation}
V_{3b}(\rho) = v_{3b} \exp(-\rho/\rho_{3b})^2.
\label{eq:3bforce}
\end{equation}
Here, the range parameter is fixed to $\rho_{3b}=5$ fm, and the strengths $v_{3b}$ depend on $j^\pi$.

\subsection{The $^{17}$Ne ground state}
\begin{figure}
\centering
\includegraphics[width=.85\linewidth]{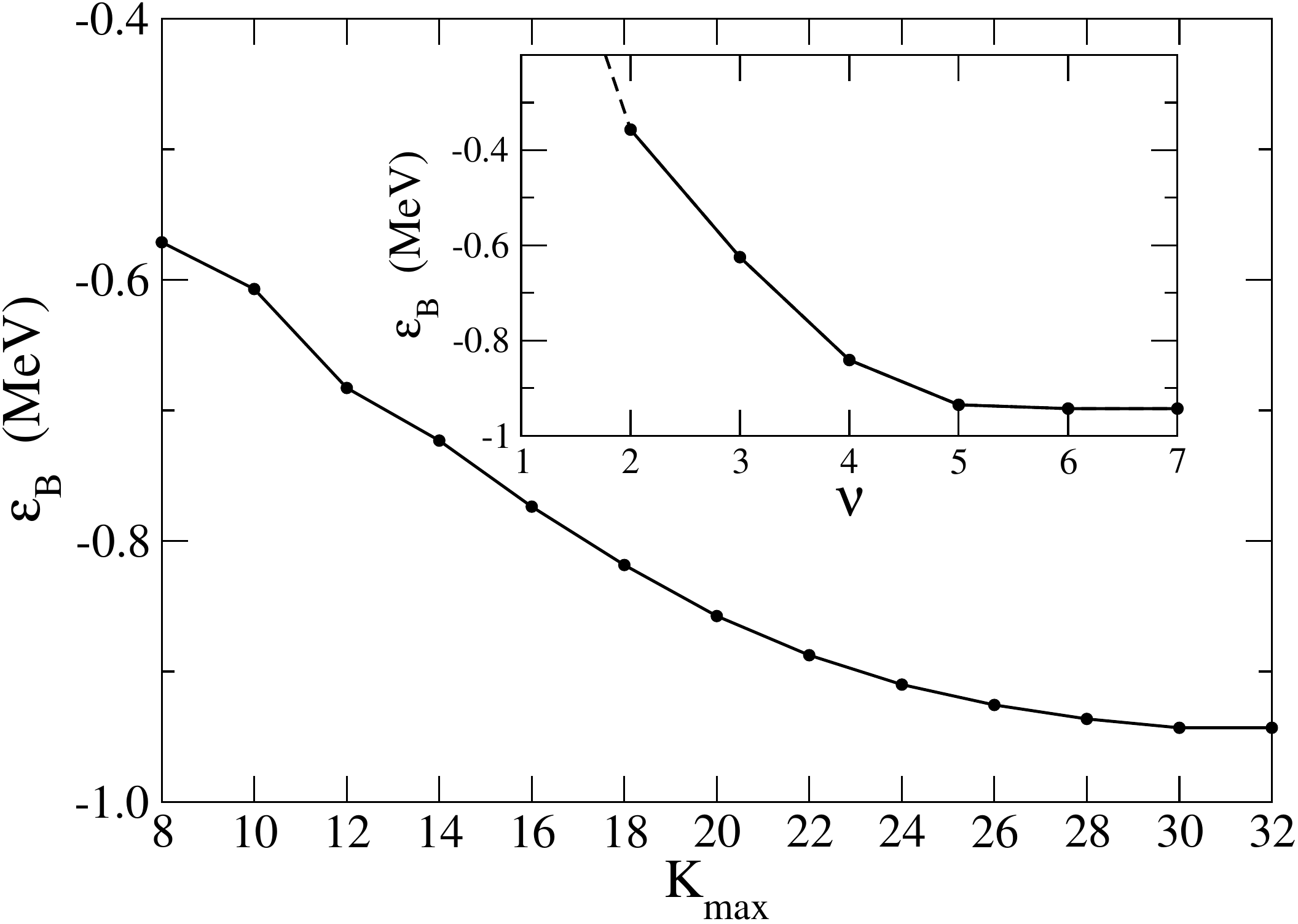}
\caption{Convergence of the ground-state energy of $^{17}$Ne with respect to the maximum hypermomentum $K_{\rm max}$ in the THO method and in terms of the number of adiabatic channels $\nu$ included in the calculation within the HA method (inset).}
\label{fig:ebound17Ne}
\end{figure}

\begin{figure}
\centering
\includegraphics[width=.95\linewidth]{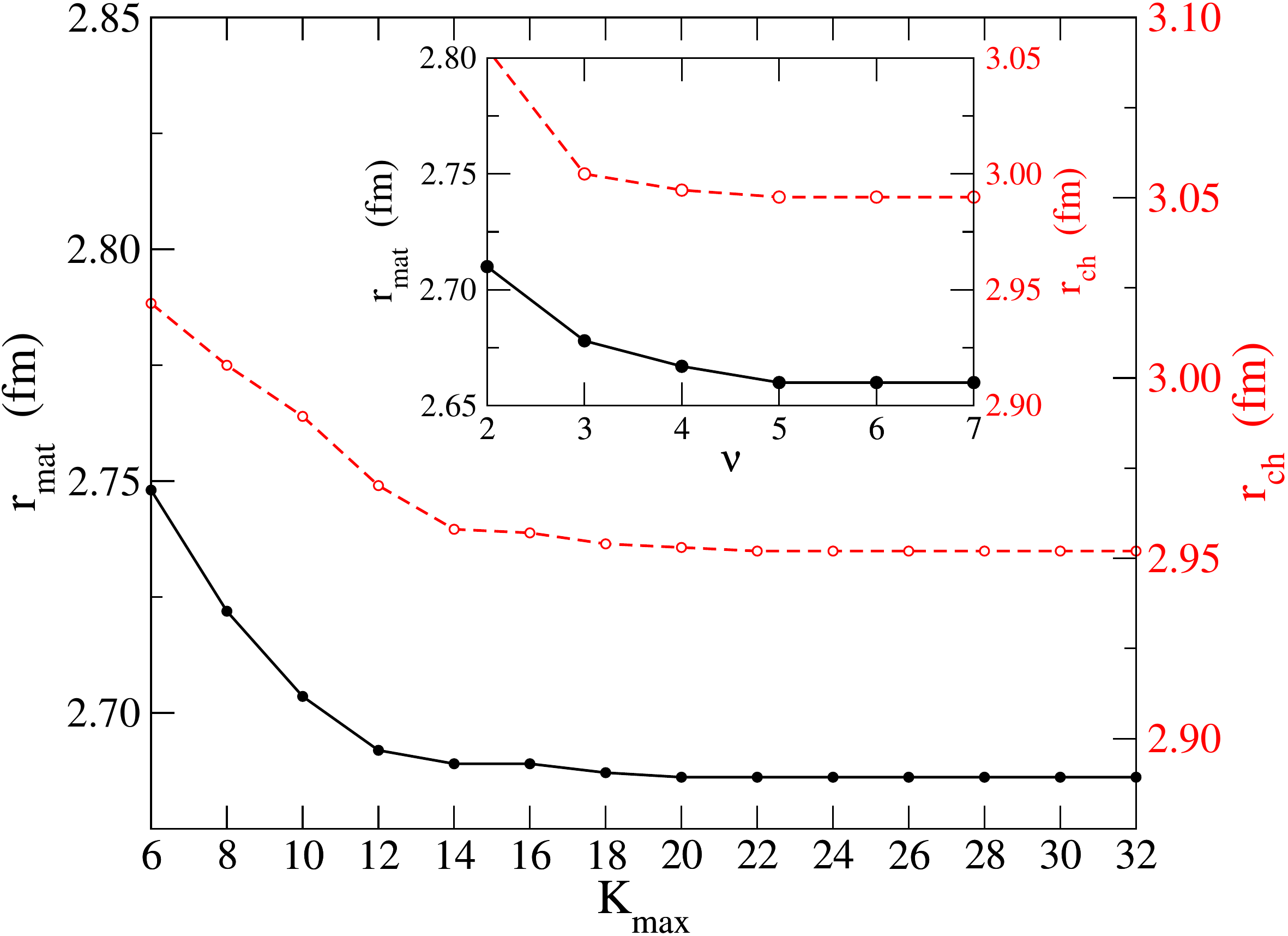}
\caption{(Color online) Convergence of the matter radius (solid line) and the charge radius (dashed red line) of $^{17}$Ne with respect to the maximum hypermomentum $K_{\rm max}$ in the THO method and in terms of the number of adiabatic channels $\nu$ included in the calculation within the HA method (inset). Notice the different scales for the matter and charge radii.} 
\label{fig:radii17Ne}
\end{figure}

Within the analytical THO method we describe the 1/2$^-$ ground state of $^{17}$Ne using a basis defined by parameters $b=0.7$ fm and $\gamma = 1.4$ fm$^{1/2}$ (see Sec.~\ref{sec:THO}). The convergence of the ground state with respect to the size of the model space, given by the maximum hypermomentum $K_{max}$, is shown in Fig.~\ref{fig:ebound17Ne}, for the ground-state energy, and in Fig.~\ref{fig:radii17Ne} for the matter and charge radii. Compared to other Borromean nuclei described within the same formalism, such as $^6$He~\cite{JCasal13} or $^9$Be~\cite{JCasal14}, the convergence for $^{17}$Ne is relatively slower. This behavior is associated with the presence of three charged particles, which enhances Coulomb effects, leading to a slower convergence in the hyperspherical expansion given by Eq.~(\ref{eq:wf}). 
These calculations are performed with a fixed value of $i_{max}=20$. To achieve converged energy and radii, $K_{max}$ has to be increased up to 30. When the HA expansion method is used, the convergence of the ground
state two-proton separation energy and the matter and charge radii are shown in the inset of
Figs.~\ref{fig:ebound17Ne} and~\ref{fig:radii17Ne}, respectively, as a function of the number
of adiabatic channels $\nu$ included in the expansion (\ref{eq3}). As shown in the figures, five adiabatic
terms are enough to get convergence.

In order to fix the 1/2$^-$ ground state to the experimental energy of --0.943 MeV~\cite{Guimaraes98,Chromik02}, a three-body strength $v_{3b}=-1.94$ MeV is required in the THO case and $v_{3b}=-2.05$ MeV in the HA case. Assuming that the $^{15}$O matter and charge radii are 2.44~\cite{Ozawa01} and 2.69 fm~\cite{Angeli13}, respectively, the computed matter and charge radii of $^{17}$Ne result 2.69 and 2.95 fm, respectively, for the THO calcultation, and 2.66 and 2.99 fm, respectively, for the HA calculation. The calculated matter radius is in good agreement with the available experimental data of $r_{mat}=2.75(7)$ fm~\cite{Ozawa01}. For the charge radius, the present result slightly underestimates the experimental value of $r_{ch}=3.042(21)$ fm~\cite{Geithner08}. This could be a consequence of the approximations within the models. Nevertheless, the three-body models with the two-body interactions presented above describe the overall features of the system spatial distribution, with a charge radius being slightly larger than the matter radius, as expected for a system comprising two valence protons. 

\begin{figure}
\centering
\includegraphics[width=\linewidth]{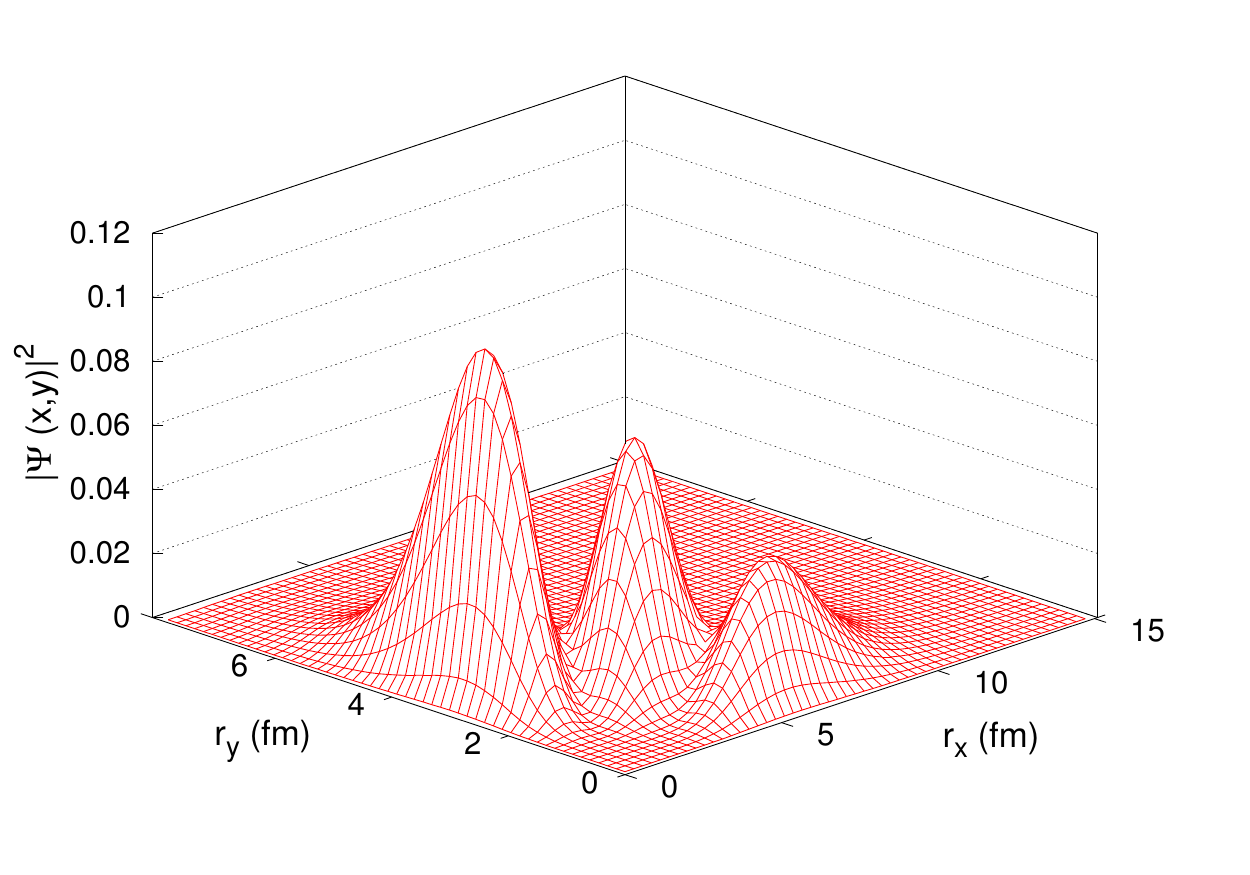}
\caption{(Color online) Probability distribution of the $^{17}$Ne ground state with the THO method.}
\label{fig:prob17Ne}
\end{figure}

The ground-state probability distribution for $^{17}$Ne in the Jacobi-$T$ set is shown in Fig.~\ref{fig:prob17Ne} for the THO calculation, where $r_x$ refers to the distance between the two valence protons. The corresponding HA result is not shown but is essentially identical. A prominent peak is observed for $r_x\simeq2.5$ fm and $r_y\simeq3$ fm. Another smaller peak is found for corresponding distances of about 5 and 1 fm, respectively. The third peak between the other two is defined by $r_x\simeq4$ fm and $r_y\simeq2$ fm. The first two peaks can be described as two protons either on the same side of the core or at almost opposite sides. The third peak accounts for intermediate configurations. This probability distribution is similar to that presented in Ref.~\cite{Garrido04}, and the differences in the relative height of the peaks are associated with the different binary $p$-$^{15}$O potentials.

The structure of $^{17}$Ne can be studied by calculating
the percentage of the total norm provided by each 
angular component $\{l_x,l_y,l,S,j_{ab}\}$. The information
about the $l$-content of the single particle proton
wave function is hindered in the Jacobi-$T$ set. A rotation
to the Jacobi-$Y$ set, where $x$ connects the $^{15}$O core
and one proton, can be performed. This transformation
is developed in Ref.~\cite{IJThompson04} and is related to the Reynal-
Revai coefficients~\cite{RR70}. The results indicate that $d$ waves
contribute with roughly 63--62\% (from THO and HA),
while 30--31\% (from THO and HA, respectively) of the
norm comes from $s$ waves.

The debate about the halo structure of $^{17}$Ne is
still unresolved. In fact, the answer to this question is, to 
a large extent, determined by the concept of halo itself.
The presence of a dilute tail in the density and charge
distributions~\cite{Tanaka10,Geithner08} led the authors to assign a halo
character to these tails, similar to what was done with
well-established neutron halo nuclei. This conclusion
was, however, questioned in Ref.~\cite{Warner98}, where the width of the 
momentum distributions after fragmentation of $^{17}$Ne
and the two-proton removal cross sections led the authors
to argue against the existence of a halo in $^{17}$Ne. 
Furthermore, from the point of view of a halo as a tunneling phenomenon,
where the nucleons in the halo reside mostly in the classically
forbidden region, it is quite clear that $^{17}$Ne can not
be considered as a quantum halo system~\cite{Garrido04,Jensen03}, 
even if its structure can be well described as a three-body system.

\subsection{$^{15}\text{O}(2p,\gamma){^{17}\text{Ne}}$ reaction rate}\label{sec:rate}
To compute the two-proton capture reaction on $^{15}$O to produce $^{17}$Ne, the electromagnetic transition probability distributions between the 1/2$^-$ ground state and $j^\pi$ continuum states are required. Previous works~\cite{Grigorenko05,Grigorenko06} suggested that for a broad range of temperatures, the rate was dominated by non-resonant $E1$ contributions, with resonant capture being relevant around 0.1--1 GK only. 

We evaluate the dominant $E1$ contribution from 1/2$^+$ and 3/2$^+$ states. As Eq.~(\ref{eq:rate1}) makes no assumption about the reaction mechanism, our approach includes sequential and direct, resonant and non-resonant contributions on an equal footing. In the THO method the continuum states are computed, for each $j^\pi$, in an analytical THO basis defined by parameters $b=0.7$ fm and $\gamma=1.0$ fm$^{1/2}$. This produces a larger level density near the breakup threshold and allows us to map the low-energy continuum with detail. Calculations are performed with $K_{max}=30$ and $i_{max}=40$ hyperradial excitations in each channel. Within the HA expansion method the $j^\pi$ continuum states are obtained after discretization of the spectrum by imposing a box boundary condition at $\rho_{max}=400$ fm. Five adiabatic terms are included in the expansion (\ref{eq3}), which are enough to get convergence in the results presented here.
The position of the low-energy 1/2$^+$ resonance at 0.96 MeV above the three-body threshold~\cite{Chromik02} will play a relevant role, and we fix its energy using $v_{3b}=-6.75$ MeV in Eq.~(\ref{eq:3bforce}), in the THO method, and $v_{3b}=-7.20$ MeV in the HA method. In the case of 3/2$^+$ states, the presence of a resonance around 3 MeV has been suggested~\cite{Guimaraes98}. Only the states close to the breakup threshold will be crucial for the reaction rate, and therefore no three-body force is included for the computation of 3/2$^+$ states. 

The electric dipolar transition probabilities between the 1/2$^-$ ground state and 1/2$^+$, 3/2$^+$ continuum states are calculated with Eq.~(\ref{eq:BE}). Using Eqs.~(\ref{eq:Qop}) and~(\ref{eq:harmonicp}), the electric dipolar operator for a system comprising two identical protons and a charged core, such as $^{17}$Ne, can be written in the Jacobi-$T$ set as
\begin{equation}
 \widehat{\mathcal{O}}_{1 M_{1}}=\mathcal{A}\left(\frac{4\pi}{3}\right)^{1/2}yY_{1M_1}(\widehat{y}),
 \label{eq:Qop3noid}
\end{equation}
and the corresponding sum rule for dipolar transitions is
\begin{equation}
\begin{split}
 S_{T}(E1) & = \sum_{nj}B(E1)_{n_0j_0\to nj} \\ & = \mathcal{A}^2\frac{3}{4\pi} \langle n_0j_0\mu_0|y^2|n_0j_0\mu_0\rangle.
\end{split}
\label{eq:sumrule1}
\end{equation}
Here, $|n_0j_0\mu_0\rangle$ represents the ground state, and the constant $\mathcal{A}$ can be easily obtained from Eqs.~(\ref{eq:relpos}) and~(\ref{eq:harmonicp}). The computation of the transition probability matrix elements provides a set of discrete values. The sum over $B(E1)$ discrete values for transitions to 1/2$^+$ states up to 15 MeV is 0.545 e$^2$fm$^2$ with the THO method and 0.510 e$^2$fm$^2$ with the HA expansion method. This, together with transitions to 3/2$^+$ states at higher energies, converges rapidly to the result provided by the sum rule in Eq.~(\ref{eq:sumrule1}), 1.687 e$^2$fm$^2$.

\begin{figure}
\centering
\includegraphics[width=0.85\linewidth]{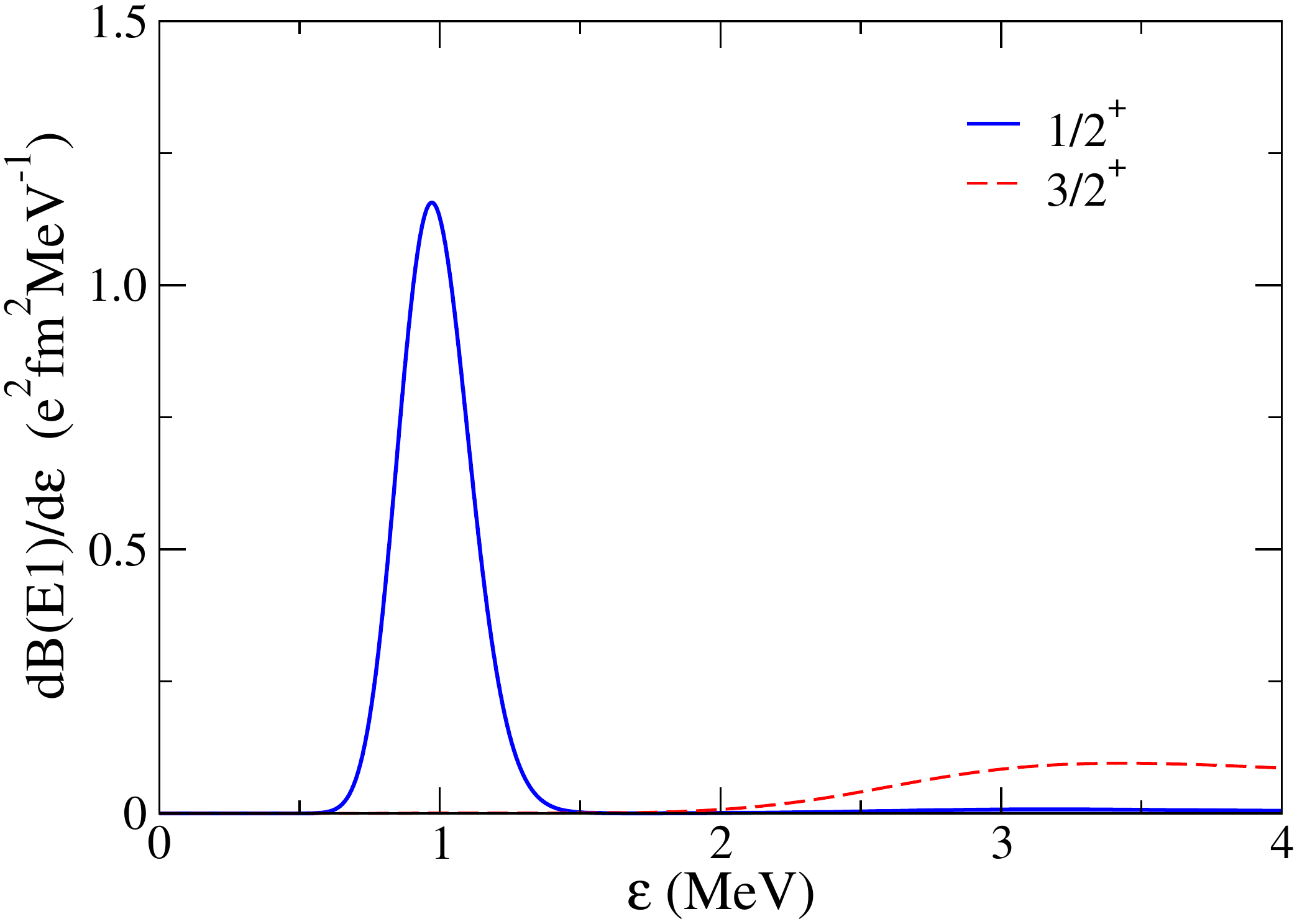}
\caption{(Color online) $B(E1)$ transition probability distribution from the 1/2$^-$ ground state to 1/2$^+$ (blue line) and 3/2$^+$ (dashed red line) continuum states in $^{17}$Ne using the THO method.}
\label{fig:dBE1ne17}
\end{figure}

\begin{figure}
\centering
\includegraphics[width=0.85\linewidth]{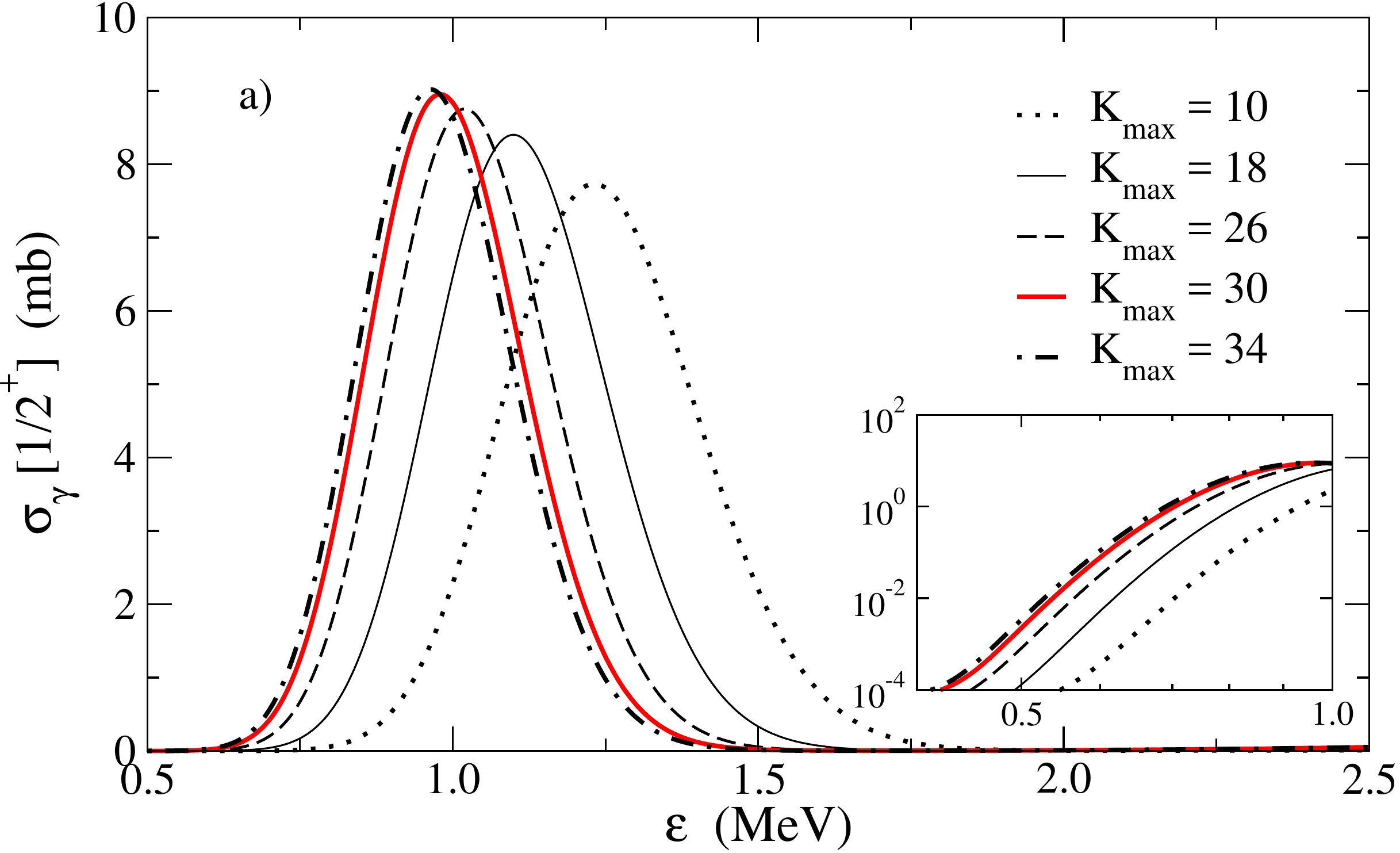}
\includegraphics[width=0.85\linewidth]{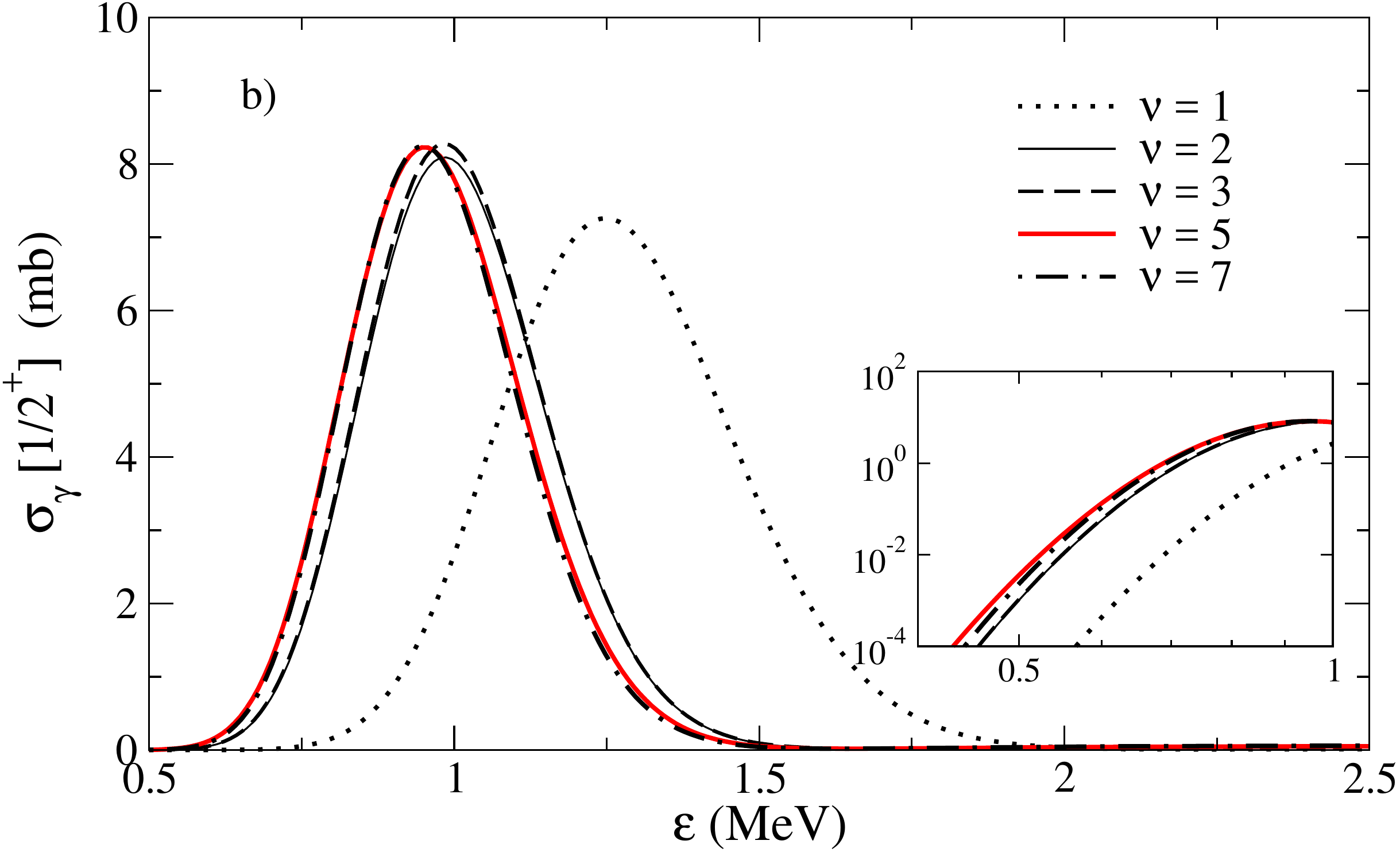}
\caption{(Color online) Convergence of the 1/2$^+$ contribution to the photodissociation cross section with respect to (a) $K_{max}$ in the THO method and (b) $\nu$ in the HA method. The inset shows the low-energy region in logarithmic scale, in order to confirm the convergence as the excitation energy reaches the threshold.}
\label{fig:E1conv}
\end{figure}

In order to obtain a continuous distribution from the discrete values, we follow the prescription presented in Ref.~\cite{JCasal14}, using Poisson distributions as smoothing functions. The present results are shown in Fig.~\ref{fig:dBE1ne17} for 1/2$^+$ states (blue line) and 3/2$^+$ states (dashed red line), using the THO method. From this figure, it is clear that a significant part of the $E1$ strength goes to the 1/2$^+$ resonance. 
The convergence of these calculations with respect to the size of the model space is shown in Fig.~\ref{fig:E1conv}(a), where the 1/2$^+$ contribution to the photodissociation cross section obtained for different values of $K_{max}$ is shown. It is clear that the calculations with $K_{max}=30$ and $K_{max}=34$ are very close together. This indicates it is safe to fix $K_{max}=30$ and adjust the position of the 1/2$^+$ resonance using the three-body force, provided the shape of the distribution is unaffected.

\begin{figure}
\centering
\includegraphics[width=0.85\linewidth]{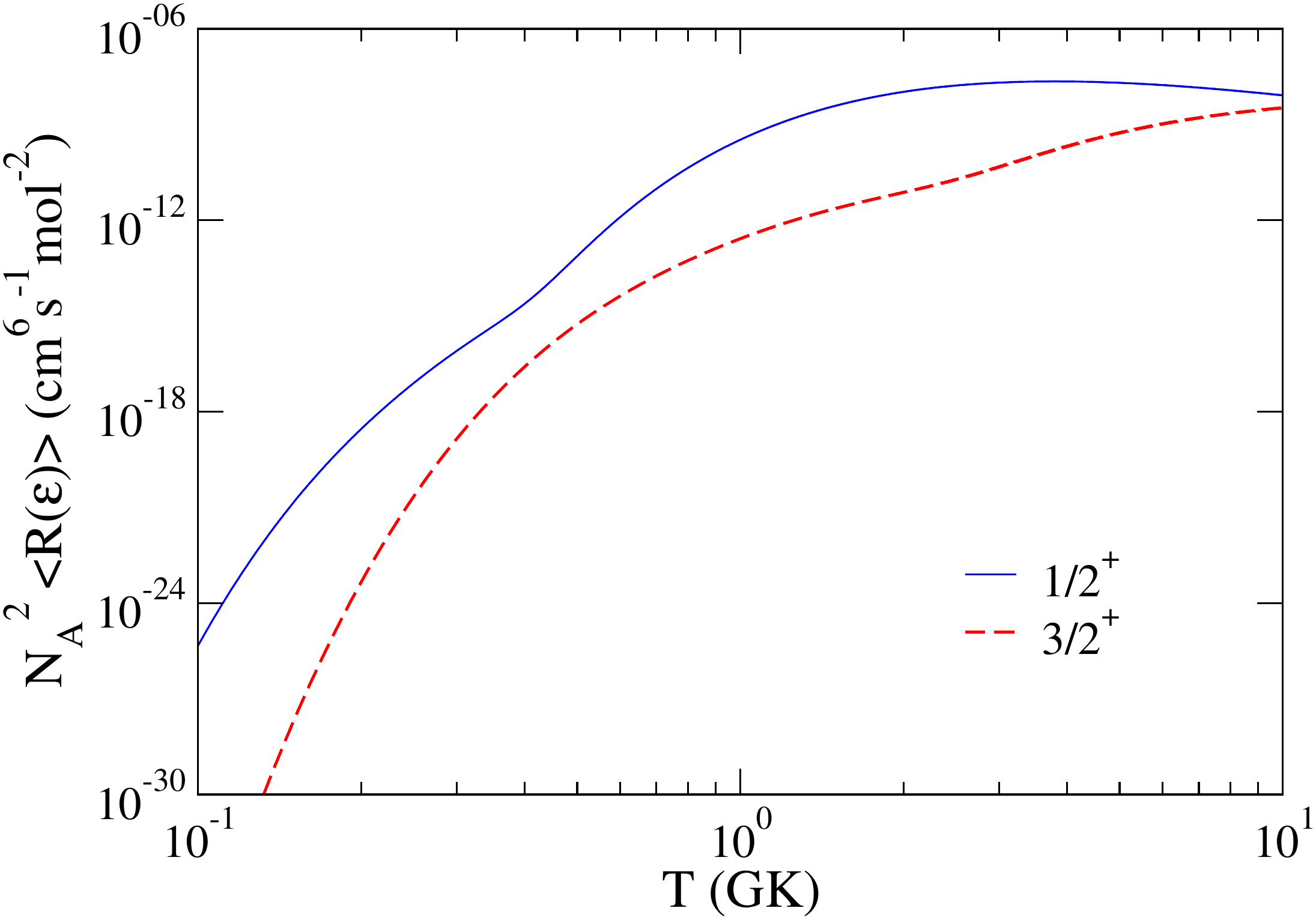}
\caption{(Color online) Reaction rate for $^{17}$Ne formation as a function of the temperature in GK, using the THO method. The two $E1$ contributions from 1/2$^+$ (blue line) and 3/2$^+$ states (dashed red line) are shown.}
\label{fig:ratej}
\end{figure}

The relationship between the transition probability distribution for dipolar transitions and the corresponding radiative capture reaction rate is given by Eqs.~(\ref{eq:rate1}) and~(\ref{eq:xsection}). The E1 contributions to the $^{15}\text{O}(2p,\gamma){^{17}\text{Ne}}$ reaction rate from 1/2$^+$ (blue line) and 3/2$^+$ states (dashed red line) using the THO method are shown in Fig.~\ref{fig:ratej} as a function of the temperature in GK. As expected, 1/2$^+$ states dominate the reaction rate in the whole temperature range. 

\begin{figure}
\centering
 \includegraphics[width=0.85\linewidth]{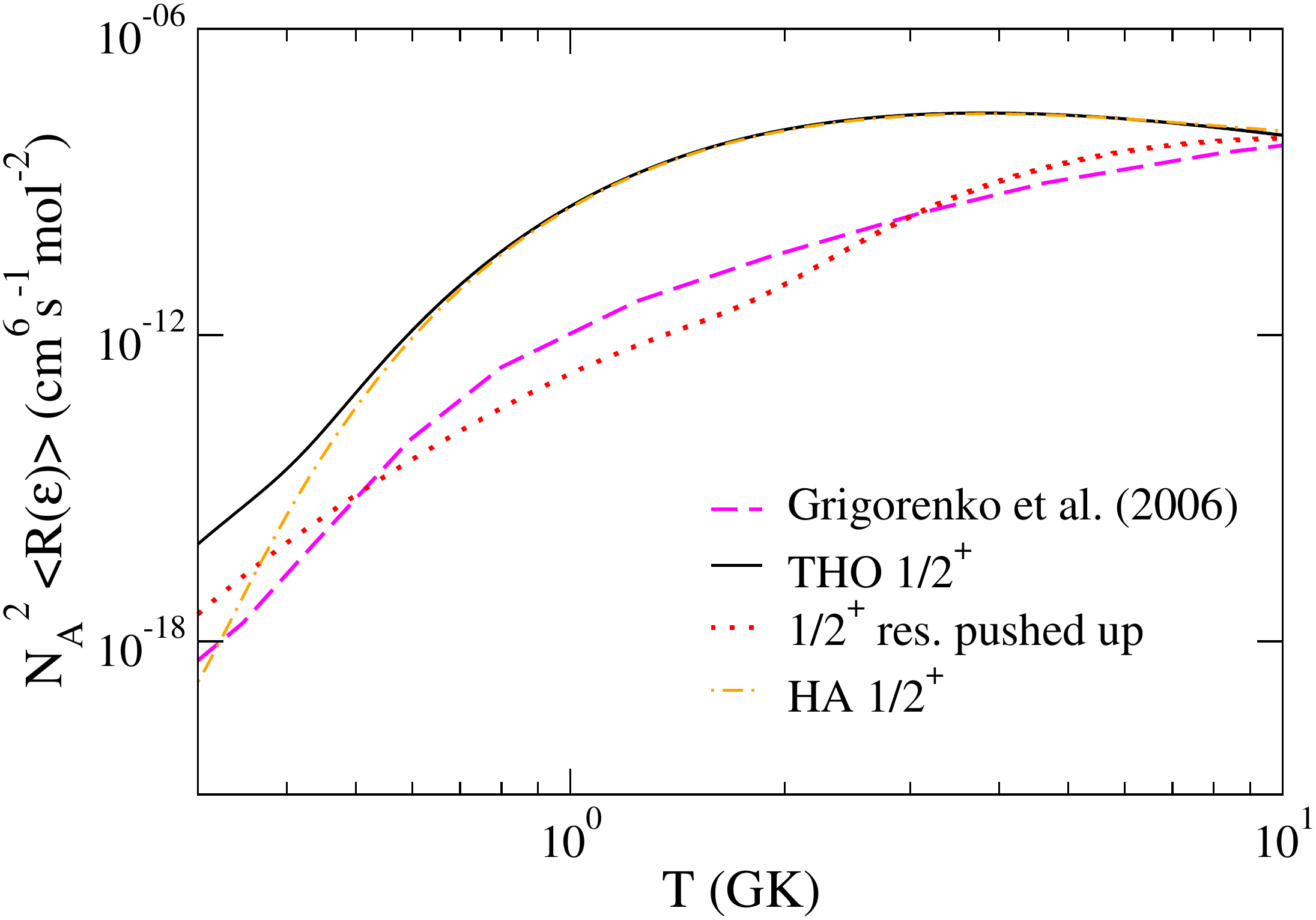}
  \caption{(Color online) Contribution to the $^{15}\text{O}(2p,\gamma){^{17}\text{Ne}}$ reaction rate from 1/2$^+$ states compared with the results in Ref.~\cite{Grigorenko06} (pink dashed line). The three-body calculations using the THO method (black solid) and the HA (dot-dashed orange) are presented. A calculation with the 1/2$^+$ resonance pushed up to higher energies is also shown (red dotted line; see the text for details).}
 \label{fig:rate_comp}
\end{figure}

The $E1$ contribution to the reaction rate from 1/2$^+$ states within the THO method is compared in Fig.~\ref{fig:rate_comp} (black solid line) with the previous calculation by Grigorenko \textit{et al.}~\cite{Grigorenko06} (pink dashed), which considers the resonant and non-resonant contributions separately. As seen in the figure, the calculation in the present work is orders of magnitude larger than the total rate 
given in Ref.~\cite{Grigorenko06}.  The value of the reaction rate at large temperatures is essentially 
determined by the $E1$ transition between the 1/2$+$ resonance
(at 0.96 MeV) and the 1/2$-$ ground state. In this energy region
the calculations of the crucial continuum states in Ref.~\cite{Grigorenko06} are not 
true three-body calculations, since the dipole states are apparently
constructed as the core-proton two-body resonance combined with
some $p$ interaction between the two-body system and the second
proton. This cannot account for genuine three-body resonances. The consequence of this is that the $dB/d\varepsilon$ strength function in Ref.~\cite{Grigorenko06} is very different to ours, with
a peak at a clearly higher energy ($\sim$4 MeV) than what it should be 
according to the experimental value of the 1/2$^+$ resonance ($\sim$1 MeV). The authors of Ref.~\cite{Grigorenko06} claim that their calculations 
include only non-resonant contributions, while the resonant part corresponding to the 1/2$^+$ is estimated in Ref.~\cite{Grigorenko05}. However, in Ref.~\cite{Grigorenko05}  
it is not clear which amount of the $E1$ strength is covered by transitions to 1/2$^+$ states, and therefore an immediate comparison with 
our present results is not possible. In order to understand the differences, we have performed a new calculation setting the three-body force to zero, which moves the 1/2$^+$ resonance energy up to $\sim 2.7$ MeV.
When this is done, we obtain the dotted curve shown in Fig.~\ref{fig:rate_comp}, which resembles much better the result given in Ref.~\cite{Grigorenko06}. This might indicate that a possible explanation for the different reaction rates could be that, in Ref.~\cite{Grigorenko05,Grigorenko06}, either the 1/2+ resonance is not located at the known experimental energy or the $E1$ strength to this resonance is not properly accounted for. It is in fact remarkable that just the $1/2^+$ contribution in the present work is noticeably larger than the total rate by Grigorenko \textit{et al.} at high temperatures.

To assess the validity of THO results, we also include in Fig.~\ref{fig:rate_comp} the 1/2$^+$ contribution within the HA method (dot-dashed orange line). This is obtained from the corresponding $B(E1)$ distribution following the same smoothing procedure as in the THO results. For consistency, in Fig.~\ref{fig:E1conv}b, the convergence of the corresponding 1/2$^+$ contribution to the photodissociation cross section within the HA calculations in terms of the adiabatic terms $\nu$ is also presented. As seen in the figure, $\nu=5$ is enough to get a sufficient convergence in the photodissociation cross section. This ensures that both approaches provide robust numerical results and can be compared properly. 

 
The temperature range of astrophysical interest in novae and x-ray bursts, where the reaction $^{15}\text{O}(2p,\gamma){^{17}\text{Ne}}$ may play a role, is of the order of 0.3--3 GK (see, for instance, Ref.~\cite{Wiescher99}). In this range, both approaches, using the analytical THO method and the HA method, agree reasonably and provide a reaction rate several orders of magnitude larger than that in Ref.~\cite{Grigorenko06}. This could imply important differences in the temperature-density profile that determines the conditions for the ${^{15}\text{O}}(2p,\gamma){^{17}\text{Ne}}$ reaction to be relevant for the rp-process. The two present calculations show differences only at very low temperatures.
We believe this discrepancy is related to the different discretization
methods used in both approaches. In particular, as mentioned in Sec.~\ref{sec:rate}, 
in the HA method the continuum spectrum is discretized by imposing a box
boundary condition with a box size of $\rho_{max}=400$ fm. When doing so
the density of states at low energies is very likely too low. The separation
between two consecutive discrete states goes like $1/\rho_{max}$, which implies
that a substantial increase of the density of states at low energies requires
a box which is too big to be implemented numerically. On the contrary,
in the THO method, the local scale transformation can be set such that a large amount 
of states concentrate at low energies. A detailed analysis of the low-energy
behavior of the reaction rate will be made in a future work. In any case,
it is clear from Fig.~\ref{fig:rate_comp} that the previous estimation in Ref.~\cite{Grigorenko06} is inconsistent with the present calculations within both the THO and the HA methods. Full rp-process network 
calculations are asked for to test the sensitivity of the trigger conditions 
of x-ray bursts to the present reaction rates.

\section{Summary and conclusions}
\label{sec:conclusions}
The structure of the Borromean nucleus $^{17}$Ne ($^{15}\text{O}+p+p$) has been described in a full three-body model using two different discretization procedures: the analytical THO method and the HA expansion method. Using the same binary potentials between the interacting pairs, both approaches provide consistent results in describing the overall features of the $^{17}$Ne ground state.

The rate of the two-proton capture reaction on $^{15}$O to produce $^{17}$Ne is computed from the $E1$ probability distributions between the 1/2$^-$ ground state and 1/2$^+$, 3/2$^+$ continuum states. The present model makes no assumption about the reaction mechanism, thus including the resonant and non-resonant, direct and sequential contributions on an equal footing. It is found that the reaction rate obtained within the THO and HA methods agree in a broad range of temperatures and provide results several orders of magnitude larger than the only previous theoretical estimation by Grigorenko \textit{et al}. This large difference could have implications for the rp-process in type I x-ray bursts and should be further investigated. 
\vspace{-10pt}

\begin{acknowledgments}
This work has been partially supported by the Spanish Ministerio de Econom\'ia y Competitividad under Projects No.~FIS2014-53448-c2-1-P, No.~FIS2014-51941-P, and No.~FIS2014-51971-P, and by Junta de Andaluc\'ia under Group No.~FQM-160 and Project No.~P11-FQM-7632. J.~Casal acknowledges support from the Ministerio de Educaci\'on, Cultura y Deporte, FPU
Research Grant No.~AP2010-3124. M.~Rodr\'iguez-Gallardo acknowledges postdoctoral support from the Universidad de
Sevilla under the V Plan Propio de Investigaci\'on, Contract No.~USE-11206-M. R.~de Diego acknowledges support by the Fundação para a Ciência e a Tecnologia (FCT) through Grant No. SFRH/BPD/78606/2011.
\end{acknowledgments}

\newpage 

\bibliography{bibfile}

\end{document}